\documentclass[]{aipproc}
\layoutstyle{6x9}

\usepackage{times,colordvi,amsmath,epsfig,float,color,multicol,subfigure,natbib}
\usepackage[latin1]{inputenc}

\newcommand\hete{{\it HETE}}
\def\arcmin{$^{\prime}$}
\newcommand{\aaps}{{\it A\&AS}}
\newcommand{\apj}{{\it ApJ}}
\newcommand{\gcn}{{\it GCN}}
\def\arcsec{$^{\prime\prime}$}
\def\chandra{{\it Chandra}} 
\def\lessim{\mathrel{\hbox{\rlap{\hbox{\lower4pt\hbox{$\sim$}}}\hbox{$<$}}}}

\begin{document}

\title 
      [Observations of XRF~030723]
      {Optical and X-ray Observations of the Afterglow to XRF~030723}

\author{N. Butler}{
  address={Center for Space Research, Massachusetts Institute of Technology, MA},
}

\iftrue
\author{A. Dullighan}{
  address={Center for Space Research, Massachusetts Institute of Technology, MA},
}
\author{P. Ford}{
  address={Center for Space Research, Massachusetts Institute of Technology, MA},
}
\author{G. Ricker}{
  address={Center for Space Research, Massachusetts Institute of Technology, MA},
}
\author{R. Vanderspek}{
  address={Center for Space Research, Massachusetts Institute of Technology, MA},
}

\author{K. Hurley}{
  address={Space Sciences Laboratory, Berkeley, CA},
}
\author{J. Jernigan}{
  address={Space Sciences Laboratory, Berkeley, CA},
}

\author{D. Lamb}{
  address={Department of Astronomy, University of Chicago, IL},
}
\author{C. Graziani}{
  address={Department of Astronomy, University of Chicago, IL},
}

\fi

\copyrightyear  {2003}

\begin{abstract}
The X-ray-flash XRF~030723 was detected by the HETE satellite
and rapidly disseminated, allowing for an optical transient to be detected
$\sim 1$ day after the burst.  We discuss observations in the
optical with Magellan, which confirmed the fade of the optical
transient.
In a 2-epoch ToO observation with Chandra, we discovered a fading 
X-ray source spatially coincident with the optical transient.
We present spectral fits to the X-ray data.  We also discuss the
possibility that the source underwent a rebrightening in the
X-rays, as was observed in the optical.  We find that the significance
of a possible rebrightening is very low ($\sim 1 \sigma$).
\end{abstract}

\maketitle

\section{Observations}
\label{sec:030528_dark}

The X-ray-flash (i.e. for the fluence $S$, 
$\log [S_X(2-30~{\rm kev})/S_\gamma(30-400~{\rm kev})]
> 0.0$)
XRF~030723 was detected by the \hete~satellite
\citep{prig03} with a 2\arcmin~radius (90\% confidence)
SXC localization.
An optical transient (OT) was reported
approximately three days after the burst by \citet{fox03}.
These authors observed a fade from $R\sim 21.3$ by 1.1 mag between
1.23 and 2.23 days after the burst.

On 25 July 2003, the {\it Chandra Observatory} targeted the field of 
XRF~030723 for a 25 ksec (E1) observation spanning the
interval 09:52-17:05 UT on 25 July, 51.4 - 59.0 hours after the 
burst. The SXC error circle from \citet{prig03} was completely 
contained within the field-of-view of the \chandra~ACIS-I array.
On 4 August 2003, \chandra~re-targeted the field of 
XRF~030723 for an 85 ksec followup (E2) observation, spanning the interval 4
August 22:22 UT to 5 August 22:27 UT, 12.69 to 13.67 days after the
burst.  For this observation, the SXC error circle from \citet{prig03}
was completely contained within the field-of-view of the \chandra~ACIS-S3 chip.

From 24.8 hours to 25.2 hours after the burst (centered on July 24.31 UT),
we observed the SXC error circle with the LDSS2
instrument on the 6.5m Magellan Clay telescope at Las Campanas
Observatory in Chile.  Four 6-minute Harris R-band exposures were
taken in $\sim 0.6$\arcsec~seeing.  Coaddition of the images gives a
limiting magnitude of $R = 24.5$.  On July 28.385 UT, 5.13 days after the
burst, we again observed the the SXC error circle with Magellan.
We obtained two 200-second exposures with the MagIC instrument in
$\sim 0.8$\arcsec~seeing, reaching a limiting magnitude of $R=24.3$.

\section{Chandra Detection and Fits}

As reported in \citet{butler03a}, 3 candidate sources
were detected within the revised SXC error region in our E1 observation.
Positions and other
data for these sources are shown in Table \ref{table:counts_030723}.
None of the sources
were anomalously bright relative to objects in \chandra~deep field
observations \citep[see, e.g.,][]{rosati02}.
The brightest object within the SXC error circle (source \#1),
lies 62\arcsec~from the center of the SXC error circle, and is within 
0.7\arcsec~of the optical transient reported by \citet{fox03}.

Table \ref{table:counts_030723} shows the number of counts detected in 
E1 and in E2.  The E2 observations were reported in \citet{butler03b}. 
Accounting for the difference in exposure times and sensitivity,
the number of counts detected for a steady source in E2 should be
$\sim 6$ times the number of counts detected in E1.  Thus, sources
3 and 4 appear to have remained constant, while source 1 has faded.
The number of counts detected in E2 corresponds to a $\sim 7\sigma$ 
significance decrease (i.e.  factor of ~6) in flux since the E1 observation. 

\begin{table}[t]
\centering
\begin{tabular}{rlrr}
\hline
\# &    Chandra Name   &    E1 Cts (bg) & E2 Cts (bg) \\
\hline
1 & CXOU J214924.4-274248  & 78.5 (1.5) & 75.6 (2.4) \\
3 & CXOU J214926.9-274146  & 19.9 (3.1) & 121.8 (4.2) \\
4 & CXOU J214928.7-274211  & 16.2 (3.8) & 98.1 (4.9) \\
\hline
\end{tabular}
\caption{
\small
Source (``Cts'') and background (``bg'') counts and positions for the 
three \chandra~sources detected within the SXC error region.  We estimate 
a position uncertainty of 1.4\arcsec. 
Astrometry was performed using six stars from the USNO-A2 catalog.
}
\label{table:counts_030723}
\end{table}

\begin{figure}[ht]
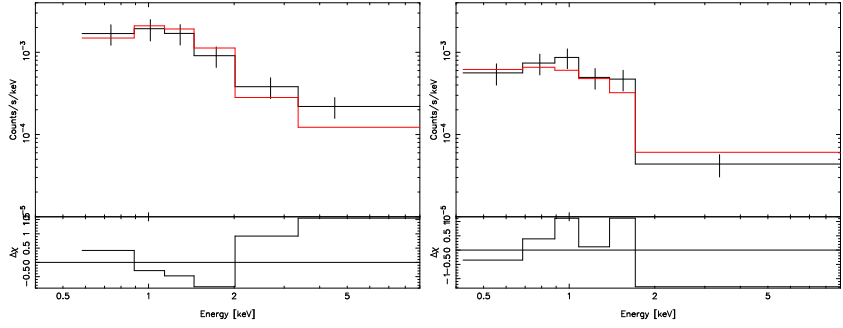

\vspace{0.5in}
\centering
\rotatebox{270}{\resizebox{10pc}{!}{\includegraphics{butler1_fig0.ps}}}
\rotatebox{270}{\resizebox{10pc}{!}{\includegraphics{butler1_fig1.ps}}}
\caption{
\small
The counts in Epoch 1 (ACIS-I, left plot) and Epoch 2 (ACIS-S, right plot) are fitted simultaneously using an
absorbed power-law model ($\chi^2_{\nu} = 8.9/9$).
}
\label{fig:030723_sp}
\end{figure}

To properly determine the fade factor we fit the E1 and E2 spectral
data jointly.  We reduce the spectral data using the standard 
CIAO\footnote{http://cxc.harvard.edu/ciao/} processing tools.
We use 
``contamarf''\footnote{http://space.mit.edu/CXC/analysis/ACIS\_Contam/script.htm
l} to correct for the quantum efficiency degradation due to
contamination in the ACIS chips, important for energies below $\sim 1$ keV.
We bin the data into 12 bins, each containing 12 or more counts,
and we fit an absorbed power-law model by minimizing $\chi^2$.  The
model has three parameters: two normalizations, and one photon index
$\Gamma$.  The absorbing column has been fixed at the Galactic value
in the source direction, $N_H = 2.4 \times 10^{20}$ cm$^{-2}$.
The model fits the data well ($\chi^2_{\nu} = 8.9/9$, 
Figure \ref{fig:030723_sp}).  The best fit photon number
index is $\Gamma = 1.9 \pm 0.3$, which is a typical
value for the X-ray afterglows of long duration GRBs \citep{costa99}.
Using this 
model, we find that the E1 flux is $(2.2 \pm 0.3) \times
10^{-14}$ erg cm$^{-2}$ s$^{-1}$ (0.5-8.0 keV band), while the E2 flux 
is $(3.5 \pm 0.5) \times 10^{-15}$ erg cm$^{-2}$ s$^{-1}$ (0.5-8.0 keV band).
The decrease in 
flux between the two epochs can be described by a power-law with a 
decay index of $\alpha= -1.0 \pm 0.1$.  This value of $\alpha$ is 
consistent with the power-law decline reported in the optical by 
\citet{dull03b} for $t\lessim 1.5$ day after the GRB; however, the 
index is considerably flatter than the index at $t>1.5$ days reported 
by \citet{dull03b}. This flatter X-ray decay may possibly be related 
to the rebrightening of the optical afterglow reported by \citet{fynbo03}.

\section{Optical Fade, Break}

\begin{figure}[ht]
\centering
\rotatebox{0}{\resizebox{15pc}{!}{\includegraphics{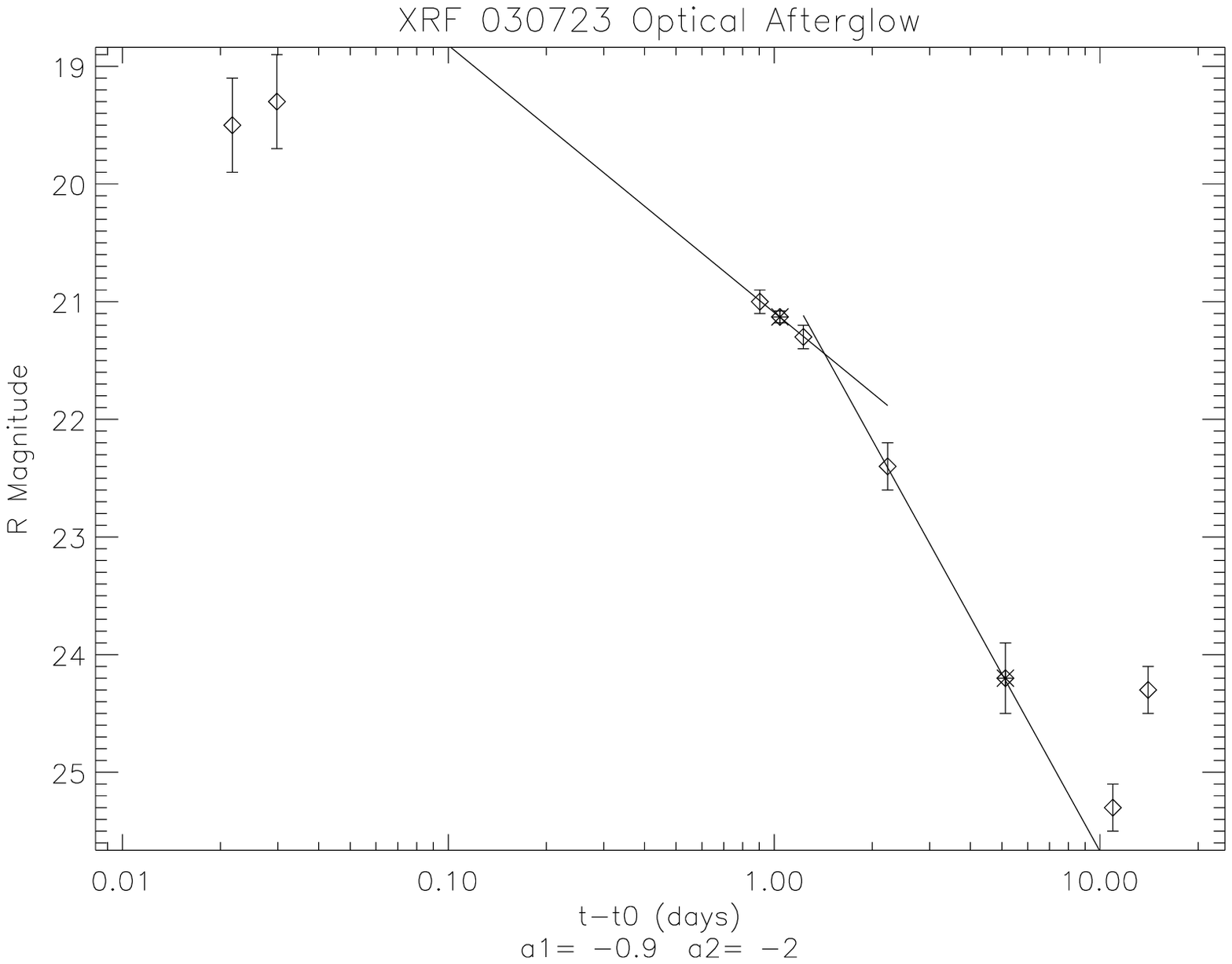}}}
\rotatebox{0}{\resizebox{15pc}{!}{\includegraphics{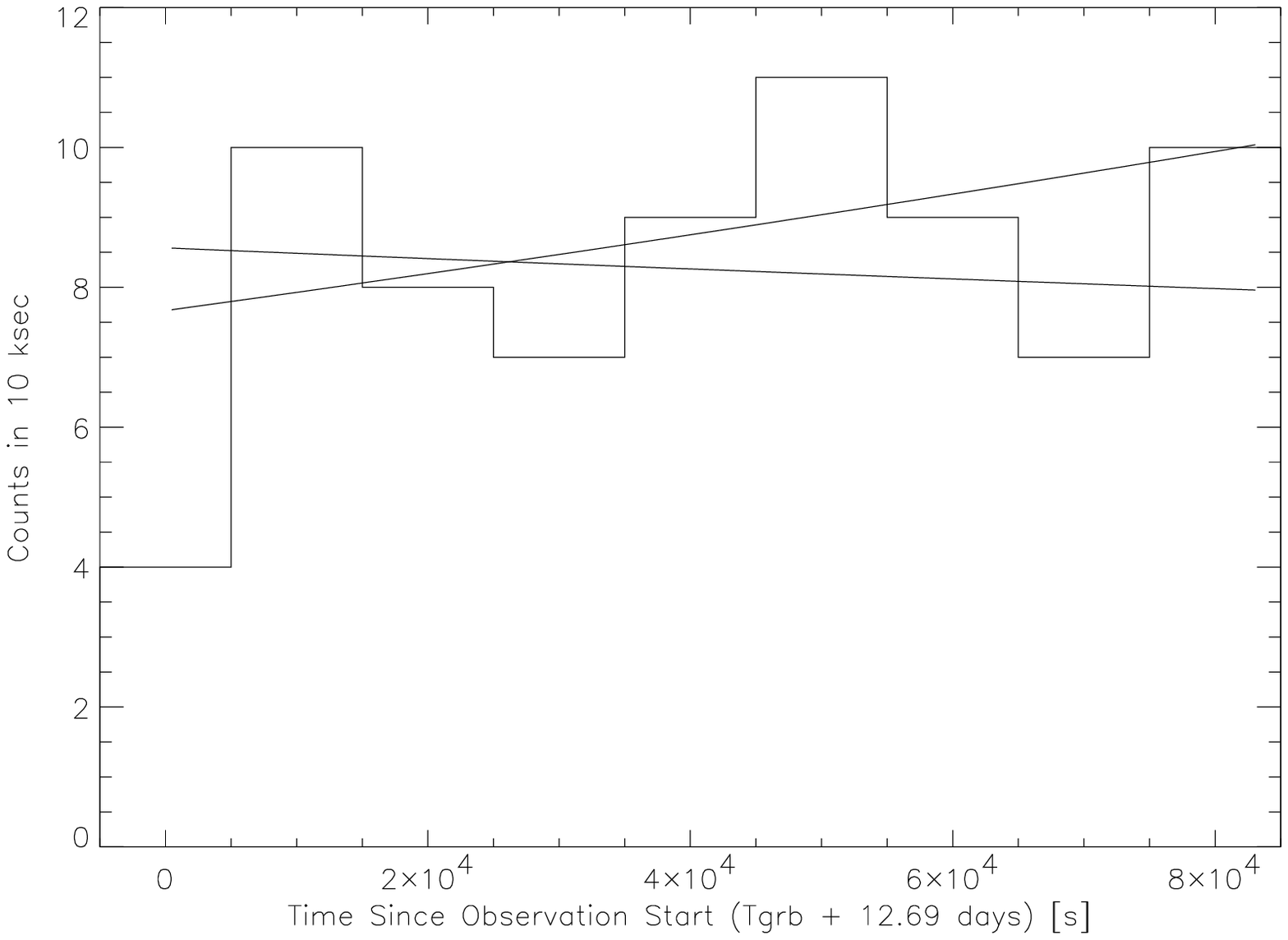}}}
\caption{
\small
Left Plot: Optical light-curve in R-band taken from reports to
the GCN \citep{fox03,dull03a,dull03b,bond03,smith03,fynbo03}.  Our data are 
marked with stars.  A temporal break may be present in the
spectrum at $t\sim 1$ day \citep{dull03b}.  The rightmost points
have been argued to imply a rebrightening \citep{fynbo03}.
Right Plot: The Count rate during the \chandra~E2 observation may be rising
as was the optical flux during the same period.  The significance
of any rise is, however, $\lessim 1\sigma$.
}
\label{fig:lc_opt}
\end{figure}

We detected the OT of \citet{fox03} in 2-epochs with Magellan,
confirming those authors claims.  Including
the other detections reported over the GCN (Figure \ref{fig:lc_opt} (a)),
we estimate a late time power law decay index of $\alpha\sim -2$, and an
early power law decay of $\alpha\sim -0.9$.  The break in the light curve 
occurs between
30-50 hours after the burst.  Our measurements have been calibrated
against the USNO photometry data reported \citet{henden03}.

\section{X-ray Rebrightening?}

The afterglow emission in the optical was apparently rebrightening during 
our E2 \chandra~observation \citep{fynbo03}.  We speculated above
that the flat decay law we measure between E1 and E2 with \chandra~may
have been in part due to a rebrightening.  To test whether this is or
is not true, we test the E2 data against two hypotheses: (1) the 
count rate versus time is described by the power-law model which
fits the overall E1,E2 {\it decay} $r(t)=8.56\times 
10^{-4}*({12.69{\rm days} \over t})$ cts/s,
(2) the count rate versus time is described by the power-law model which
fits the optical {\it rise} during the E2 observation 
$r(t)=a*({t \over 12.69{\rm days}})^{3.7} cts/s$,
where $a$ is a free parameter.  Using the arrival times $t_i$ for 75 photons,
we choose the model which maximizes the likelihood:
$$ {\cal L} (t_1,t_2,...,t_n) = 
r(t_1)\cdot r(t_2)\cdot ...\cdot r(t_n)\cdot 
\exp \left \{ -\int_{t_0}^{t_n} r(\tau) d\tau \right \}, $$
where the integral in the exponential is carried out for the good time
intervals of \chandra~data acquisition.  We find a best fit value for
$a$ of $7.67 \times 10^{-4}$ cts/s.  (Figure \ref{fig:lc_opt} (b) shows
the E2 counts in 10 ksec bins, with models (1) and (2) overplotted.)
The corresponding difference
in $\log({\cal L})$ is 0.379.  Simulating arrival times for 75 photons using
model (1), a more extreme value of $\delta \log({\cal L})$ found from
fitting both models is observed to occur for approximately 1/3 of the trials.
Thus, a rebrightening is preferred by the data, but at only $1\sigma$
significance.

\section{Conclusions}

We have derived power-law spectral parameters to X-ray data taken
in two epochs with \chandra~for XRF~030723.  The photon index $\Gamma$
we derive is a typical value for long-duration GRBs, possibly indicating
a similarity between these objects an XRFs.  The decrease in model
normalization
between the two epochs ($\Delta \chi^2 = 43.6$, for 1 additional degree
of freedom; i.e. $6.6\sigma$) confirms that source \#1 is the X-ray
counterpart to XRF~030723 and to the OT discovered by \citet{fox03}. 
Our optical observations, along with the other observations reported
over the GCN (Figure \ref{fig:lc_opt} (a)),
imply a break in the R-band light curve at $t\sim 1$ day after
the burst.  We have tested for an X-ray rebrightening, but we find 
only very weak
evidence for a rebrightening similar to that observed in the optical
by \citet{fynbo03}.

\clearpage


\noindent
\begin{thebibliography}{6}

\bibitem[Bond(2003)]{bond03}
Bond, H.~E. 2003, \gcn, 2329
\bibitem[Butler et al.(2003a)]{butler03a}
Butler, N., et al. 2003a, \gcn, 2328
\bibitem[Butler et al.(2003b)]{butler03b}
Butler, N., et al. 2003b, \gcn, 2347
\bibitem[Costa et al.(1999)]{costa99}
 Costa, E., et al. 1999, \aaps, 138, 425
\bibitem[Dullighan et al.(2003a)]{dull03a}
Dullighan, A., et al. 2003a, \gcn 2326
\bibitem[Dullighan et al.(2003b)]{dull03b}
Dullighan, A., et al. 2003b, \gcn, 2336
\bibitem[Fox et al.(2003)]{fox03}
Fox, D.~B., et al. 2003, \gcn, 2323
\bibitem[Fynbo et al.(2003)]{fynbo03}
Fynbo, J.~P.~U., et al. 2003, \gcn, 2345
\bibitem[Henden(2003)]{henden03}
 Henden, A. 2003, \gcn, 2343
\bibitem[Prigozhin et al.(2003)]{prig03}
Prigozhin, G., et al. 2003, \gcn, 2313
\bibitem[Rosati et al.(2002)]{rosati02}
 Rosati, P., et al. 2002, \apj, 566, 667
\bibitem[Smith et al.(2003)]{smith03}
 Smith, D.~A., et al. 2003, \gcn, 2338
\end{thebibliography}
\end{document}